\documentstyle[12pt]{article}
\begin{document}
\begin{flushright}
{\bf IMSc - 960101}
\end{flushright}
\vskip .5cm
\begin{center}
{\bf{FOCK SPACE REPRESENTATION OF DIFFERENTIAL CALCULUS ON \\
THE NONCOMMUTATIVE QUANTUM SPACE}}

\vskip 2cm

{\bf A.K. Mishra and G. Rajasekaran} \\
{\it Institute of Mathematical Sciences} \\
{\it C.I.T. Campus, Madras - 600 113, India} \\
{\bf e-mail : mishra@imsc.ernet.in; graj@imsc.ernet.in}

\vskip 2.5cm

{\bf ABSTRACT}
\end{center}

\baselineskip=16pt
A complete Fock space representation of the covariant differential
calculus on quantum space is constructed.  The consistency criteria for
the ensuing algebraic structure, mapping to the canonical fermions and
bosons and the consequences of the new algebra for the statistics of
quanta are analyzed and discussed.  The concept of statistical
transmutation between bosons and fermions is introduced.


\newpage

\noindent{\bf 1. Introduction}

Quantum groups, quantum vector spaces and the underlying notion of
deformations have substantially enriched the arena of mathematics and
mathematical physics.  Formulations of the covariant differential
calculi on non-commuting `quantum' spaces, and their Fock space realizations 
have recently attracted much attention$^{1-5}$.  The reinterpretation of
these differential calculi in terms of creation and annihilation
operators have led to various generalizations of Heisenberg canonical
commutation relations and enabled one to introduce particles which obey
generalized quantum statistics$^{6-8}$.

In the present communication, it is shown that the existing scheme of
mapping the differential calculus to Fock space is incomplete.  Whereas
the co-ordinates of the quantum plane are identified with creation
operators, no such identification has been made with regard to the
co-ordinates of the exterior quantum plane.  We complete this scheme by
introducing an additional set of creation and annihilation operators,
and obtain the associated commutation relations.

The Fock space corresponding to the non-commuting differential calculus
describes the states of two distinct kinds of quanta, bosonic and
fermionic.  A non-trivial consequence of the present formalism concerns
the possibility of statistical transmutation between these bosons and
fermions.  This transmutation vanishes when the deformation is removed.

In the next section, we briefly recapitulate the essentials of the
differential calculus on the quantum space.  The various steps leading
to the construction of the associated  Fock space are outlined in
Sec.3.  This is followed by Sec.4 which is on statistical transmutation.
Consistency conditions for the Fock space are discussed in Sec.5.  The
algebraic structure is completed in Sec.6  where we also give the
relationship between the new algebra and the canonical algebra of
fermions and bosons.  Sec.7 is devoted to some concluding remarks.

\vskip .5cm

\noindent {\bf{2. Differential calculus on the quantum plane}}

The ``quantum" space or plane is characterized by noncommuting
coordinates $x_i (i = 1 \ldots n)$ satisfying the q-commutation relation
:

$$
x_i x_j - q x_j x_i = 0, \quad {\rm for} \quad i < j    \eqno(1)
$$
where the deformation parameter q is a real number.  The differential
calculus in the quantum space is constructed using three sets of basic
entities, viz. (i) coordinates $x_i$, (ii) derivatives $\partial/
\partial x_i$ and (iii) differentials or coordinates of the exterior
quantum plane $dx_i$, together with the q-commutation relations among
these.  In addition to (1), the required relations are taken to be$^{1,3}$
the following set
(throughout the paper, we shall
take $i < j$, in any relation involving $i$ and $j$, unless otherwise
specified):
$$
\frac{\partial}{\partial x_i} \, \frac{\partial}{\partial x_j} -
\frac{1}{q} \frac{\partial}{\partial x_j} \, \frac{\partial}{\partial
x_i} = 0   \eqno(2)
$$
$$
\frac{\partial}{\partial x_i} x_i -q^2 x_i \frac{\partial}{\partial x_i}
= 1 + (q^2 - 1) \sum^n_{k=i+1} \, x_k \frac{\partial}{\partial x_k}
\eqno(3)
$$
$$
\frac{\partial}{\partial x_i} x_j - q x_j \frac{\partial}{\partial x_i}
= 0   \eqno(4)
$$
$$
\frac{\partial}{\partial x_j} x_i - q x_i \frac{\partial}{\partial x_j}
= 0   \eqno(5)
$$
$$
dx_i dx_j + \frac{1}{q} dx_j dx_i = 0    \eqno(6)
$$
$$
(dx_i)^2 = 0     \eqno(7)
$$
$$
x_i dx_i - q^2 dx_i x_i = 0   \eqno(8)
$$
$$
x_i dx_j - q dx_j x_i - (q^2 - 1) dx_i x_j = 0    \eqno(9)
$$
$$
x_j dx_i - q dx_i x_j  = 0    \eqno(10)
$$
$$
\frac{\partial}{\partial x_i} dx_i - \frac{1}{q^2} dx_i
\frac{\partial}{\partial x_i} - \left( \frac{1}{q^2} -1 \right)
\sum^{i-1}_{k=1} dx_k \frac{\partial}{\partial x_k} = 0     \eqno(11)
$$
$$
\frac{\partial}{\partial x_i} dx_j - \frac{1}{q} dx_j
\frac{\partial}{\partial x_i} = 0    \eqno(12)
$$
$$
\frac{\partial}{\partial x_j} dx_i - \frac{1}{q} dx_i
\frac{\partial}{\partial x_j} = 0    \eqno(13)
$$
We can also introduce the exterior differential {\it d}
$$
d = \sum_i \, dx_i \, \frac{\partial}{\partial x_i}     \eqno(14)
$$
It can be verified that, as a consequence of the relations (1-13), the
exterior differential while operating on functions f and g of the
coordinates $x_i$ satisfies Leibnitz rule :
$$
d(fg) = (df)g + f(dg)     \eqno(15)
$$
{\bf 3. Towards Fock space realization}

A partial Fock space realization of the differential calculus has been
constructed through the mapping$^{6,8,9}$ 
$$
x_i \rightarrow b^{\dagger}_i     \eqno(16)
$$
$$
\frac{\partial}{\partial x_i} \, \rightarrow b_i   \eqno(17)
$$
where $b_i$ and $b^{\dagger}_i$ are annihilation and creation operators
and one assumes the existence of a vacuum state $| 0 >$ annihilated by
all $b_i$'s:
$$
b_i | 0 > = 0    \eqno(18)
$$
These operators satisfy
the following algebra
 obtained from  (1-5):
$$
b^{\dagger}_i b^{\dagger}_j - q b^{\dagger}_j b^{\dagger}_i = 0
\eqno(1^{\prime})
$$
$$
b_i b_j - \frac{1}{q} b_j b_i = 0     \eqno(2^{\prime})
$$
$$
b_i b^{\dagger}_i -q^2 b^{\dagger}_i b_i = 1 + (q^2 - 1) \sum^n_{k=i+1}
\, b^{\dagger}_k b_k      \eqno(3^{\prime})
$$
$$
b_i b^{\dagger}_j - q b^{\dagger}_j b_i  = 0    \eqno(4^{\prime})
$$
$$
b_j b^{\dagger}_i - q b^{\dagger}_i b_j  = 0    \eqno(5^{\prime})
$$
\noindent Note that (2$^{\prime}$) and (5$^{\prime}$) are the hermitian
conjugates of (1$^{\prime}$) and (4$^{\prime}$) respectively.  In order
to have the complete Fock space realization of the differential
calculus, one has to take the following three steps.

(A) Mapping of $dx_i$ to a creation operator
$$
d x_i \longrightarrow f^{\dagger}_i    \eqno(19)
$$
Consequently, (6-13) lead to
$$
f^{\dagger}_i f^{\dagger}_j + \frac{1}{q} f^{\dagger}_j f^{\dagger}_i = 0
\eqno(6^{\prime})
$$
$$
f^{\dagger}_i f^{\dagger}_i = 0     \eqno(7^{\prime})
$$
$$
b^{\dagger}_i f^{\dagger}_i - q^2 f^{\dagger}_i b^{\dagger}_i = 0
\eqno(8^{\prime})
$$
$$
b^{\dagger}_i f^{\dagger}_j - q f^{\dagger}_j b^{\dagger}_i - (q^2 - 1)
f^{\dagger}_i b^{\dagger}_j = 0  \eqno(9^{\prime})
$$
$$
b^{\dagger}_j f^{\dagger}_i - q f^{\dagger}_i b^{\dagger}_j = 0
\eqno(10^{\prime})
$$
$$
b_i f^{\dagger}_i - \frac{1}{q^2} f^{\dagger}_i b_i - ( \frac{1}{q^2} - 1)
{\displaystyle{\sum^{i-1}_{k=1}}} f^{\dagger}_k b_k = 0
\eqno(11^{\prime})
$$
$$
b_i f^{\dagger}_j - \frac{1}{q} f^{\dagger}_j b_i = 0
\eqno(12^{\prime})
$$
$$
b_j f^{\dagger}_i - \frac{1}{q} f^{\dagger}_i b_j = 0   \eqno(13^{\prime})
$$

Thus, using the mapping relations (16,17,19), the entire algebra of
$\left\{ x, \frac{\partial}{\partial x}, dx \right\}$ given by (1 - 13)
has been converted to the algebra of $\left\{ b^{\dagger}, b,
f^{\dagger} \right\}$ given by (1$^{\prime}$ - 13$^{\prime}$).

(B) Once the creation operator $f^{\dagger}_i $ has been introduced, the
existence of the annihilation operator follows through the hermitian
conjugation, i.e. $f = (f^{\dagger})^{\dagger}$.  As an immediate
consequence, we have the following additional relations obtained by
taking hermitian conjugate of (6$^{\prime}$ - 13$^{\prime}$) :
$$
f_j f_i + \frac{1}{q} f_i f_j = 0   \eqno(20)
$$
$$
f_i f_i = 0    \eqno(21)
$$
$$
f_i b_i - q^2 b_i f_i = 0    \eqno(22)
$$
$$
f_j b_i - q b_i f_j - (q^2 - 1) b_j f_i = 0   \eqno(23)
$$
$$
f_i b_j - q b_j f_i = 0   \eqno(24)
$$
$$
f_i b^{\dagger}_i - \frac{1}{q^2} b^{\dagger}_i f_i - ( \frac{1}{q^2} -1)
{\displaystyle{\sum^{i-1}_{k=1}}} b^{\dagger}_k f_k = 0    \eqno(25)
$$
$$
f_j b^{\dagger}_i - \frac{1}{q} b^{\dagger}_i f_j  = 0  \eqno(26)
$$
$$
f_i b^{\dagger}_j - \frac{1}{q} b^{\dagger}_j f_i = 0   \eqno(27)
$$
We assume that the vacuum state is annihilated by all the $f_i$'s also
:
$$
f_i | 0 >  = 0    \eqno(28)
$$
(C) In the above eqs.(1$^{\prime}$ - 13$^{\prime}$, 20 - 27), the commutation properties between all
pairs of operators except the pair $f$ and $f^{\dagger}$ have been
specified.  In order to complete the algebraic structure for the Fock
space realization, we have to know the commutation between $f$ and
$f^{\dagger}$.  We shall achieve this final step in Sec.6.

\vskip .5cm
\noindent{\bf 4. Statistical transmutation}

In this section, we examine the nature of the Fock space generated by
the creation and annihilation operators satisfying the algebra given in
the last section.  The Fock space consists of the vacuum state $| 0 >$
defined in (18) and (28) together with the set of states obtained
by letting any product of an arbitrary number of creation operators
$b^{\dagger}_i, f^{\dagger}_j$ act on $| 0 >$.

Because of (7$^{\prime}$), we see that the f-quanta obey Pauli's exclusion
principle while there is no such restriction on the b-quanta.  Hence we
shall call the f and b as fermions and bosons respectively, although
they are not to be identified with the canonical fermions and bosons.

The algebra of the operators $b, b^{\dagger}, f$ and  $f^{\dagger}$
given in Sec.3 is not invariant
under the phase transformation:
$$
\left. \begin{array}{lclclclclclcl}
b_i & \longrightarrow & e^{i \phi_i} b_i & ; \quad b^{\dagger}_i &
\longrightarrow & e^{-i \phi_i} b^{\dagger}_i \\
&\\
f_i & \longrightarrow & e^{i \psi_i} f_i & ; \quad f^{\dagger}_i &
\longrightarrow & e^{-i \psi_i} f^{\dagger}_i 
  
\end{array} \right\}    \eqno(29)
$$
where $\phi_i$ and $\psi_i$ are arbitrary real numbers.  As a
consequence, the Fock states constructed with these operators do not
have definite values for $n_i$ and $m_i$, where $n_i$ and $m_i$ are the
number of $b_i$ and $f_i$ quanta respectively.  However, the algebra is
invariant under the above transformation (29) if $\phi_i = \psi_i$.
Hence the Fock states do have definite values for $t_i$, the total
number of quanta of index i (including bosons and fermions) : $t_i = n_i
+ m_i$.   Further, invariance of the algebra is again restored if all
$\phi_i$ and $\psi_i$ are independent of i and this implies that the
Fock states have definite values for the total number of bosons $n =
\sum_i n_i$ and the total number of fermions $m = \sum_i m_i$.
However, note that if we specify $t_i$ for all $i$ and $n$, $m$ will be
redundant, since $m = \sum_i t_i -n$.

The new feature of the Fock space which allows the bosons and fermions
to be transmuted into each other so that only the total $t_i$ can be
specified rather than $n_i$ and $m_i$ separately, can be called
{\it{statistical transmutation}}.  Actually, every such
transmutation of a boson into fermion for a particular index always
goes with the simultaneous transmuation of a fermion into boson for some
other index so that the total number of bosons and the total number of
fermions is conserved.  Hence, considering the bosons alone, one can
recognize an index transmutation also and similarly for the fermions.

Because of statistical transmutation, a multiparticle state containing
bosons and fermions has a new kind of exchange property, which can be
read off from (9$^{\prime}$).  For the state vector containing $b_i
f_j (i < j)$,  exchange of the boson $b_i$ and the fermion
$f_j$ leads to a state that is a linear superposition of the state vector
containing $f_j b_i$ with the state vector containing $f_i b_j$.

Statistical transmutation is a nontrivial complication in the construction of the new
Fock space and so one may even question the existence of such a Fock space. In
the next two sections we shall give an affirmative answer to this question.

\vskip .5cm 

\noindent{\bf 5. Consistency conditions}

Let c denote any of the annihilation operators $b_i$ or $f_i$ and
$c^{\dagger}$ denote $b^{\dagger}_i \, {\rm or} f^{\dagger}_i$.  We can
classify the algebraic relations of Sec.3, into two categories (i)
$cc^{\dagger}$ relations (3$^{\prime}$ - 5$^{\prime}$,
11$^{\prime}$ - 13$^{\prime}$, 25-27), (ii) $cc$
relations (2$^{\prime}$, 20-24).  
The $c^{\dagger} \, c^{\dagger}$ relations 
(1$^{\prime}$, 6$^{\prime}$ - 10$^{\prime}$) are just hermitian conjugates of the $cc$ relations and hence
are not independent.  It is an important fact which does not seem to be
well recognized, that within the framework of Fock space, the $cc$
relations themselves are not independent of the $cc^{\dagger}$ relations.  For,
given the vacuum state defined by eqs.(18) and (28), and the rules
for the commutation of $c$ and $c^{\dagger}$ given by the $cc^{\dagger}$
relations, {\it all} matrix elements in the Fock space can be computed.
Hence any $cc$ relation which is imposed will be either inconsistent
with the $cc^{\dagger}$ relations, or derivable from the $cc^{\dagger}$
relations, if consistent.

To derive the $cc$ relations from the $cc^{\dagger}$ relations, we
proceed as follows$^{10}$.  Let $Q^{\alpha}_{ij}$ be a quadratic in c's
such as the left-hand side of any of the $cc$ relations (2$^{\prime}$,
20-24),
with $\alpha$ denoting the equation number.  By using the $cc^{\dagger}$
algebra (3$^{\prime}$ - 5$^{\prime}$, 11$^{\prime}$ -
13$^{\prime}$, 25-27), we shall show that 
$$
Q^{\alpha}_{ij} c^{\dagger}_k = \sum_{\beta \ell mt} \, F^{\alpha \beta,
{\ell}mt}_{ijk} c^{\dagger}_t Q^{\beta}_{\ell m}     \eqno(30)
$$
\noindent where $F_{ijk}^{\alpha \beta. \ell m t}$ is some q-dependent
number. Applying this equation again, we get
$$
Q^{\alpha}_{ij} c^{\dagger}_k c^{\dagger}_p = \sum_{\beta \ell mt}
\, \sum_{\gamma uvs} \, F^{\alpha \beta, \ell mt}_{ijk} \, F^{\beta 
\gamma, uvs}_{\ell mp} c^{\dagger}_t c^{\dagger}_s \ Q^{\gamma}_{uv}
\,.  \eqno(31)
$$
\noindent Thus, $Q^{\alpha}_{ij}$ can be pushed to the right of any string  
of creation operators $c^{\dagger}_k c^{\dagger}_p \ldots $ . Also
note that the string of creation operators can contain
$b^{\dagger}$ and $f^{\dagger}$ in arbitrary order.  Allowing both sides
of equations such as (30) or (31)  to act on $| 0 >$, the right-hand-side vanishes because
of (18) and (28) and so we see that $Q^{\alpha}_{ij}$ acting on any Fock
state $c^{\dagger}_k c^{\dagger}_p \ldots | 0 >$ gives zero.  Hence we
may write the operator identity :
$$
Q^{\alpha}_{ij} = 0     \eqno(32)
$$
\noindent which are the $cc$ relations.  Thus, (30) are the necessary
and sufficient conditions for the existence of the $cc$ relations and
they are also the consistency conditions for the Fock space realization.

After a straightforward computation, we get the following results where
{\nolinebreak${\ i<j}$}.

$$\begin{array}[b]{lcl}
 Q^{2'}_{ij} b^{\dagger}_k & = & q^2 b^{\dagger}_k Q^{2'}_{ij} , \quad {\rm for}
\, k \neq i \quad {\rm or} \quad j     \\ 
 & = & q^3 b^{\dagger}_j Q^{2'}_{ij} + q (q^2 - 1) 
{\displaystyle{\sum^n_{a=j+1}}}
b^{\dagger}_a Q^{2'}_{i a}, \quad {\rm for} \quad k = j   \\   
& = & q^3 b^{\dagger}_i Q^{2'}_{ij} + q(q^2-1)
{\displaystyle{\sum^{j-1}_{a=i+1}}} \, b^{\dagger}_a Q^{2'}_{aj} 
\\
&   &  -  (q^2-1) {\displaystyle{\sum^n_{a=j+1}}} b^{\dagger}_a
Q^{2'}_{ja}, \quad {\rm for} \quad k = i      
\end{array}  \hfill \eqno(33)  $$

$$\begin{array}[b]{lcl}
Q^{2'}_{ij} f^{\dagger}_k & = & {\displaystyle{\frac{1}{q^2} f^{\dagger}_k Q^{2'}_{ij},
\quad {\rm for} \quad k \neq i \quad {\rm or} \quad j}} \nonumber \\
& = & {\displaystyle{\frac{1}{q^3} f^{\dagger}_j Q^{2'}_{ij} -  \frac{1}{q^2}( 1-q^{2}) 
\sum^{i-1}_{a = 1}}} \, f^{\dagger}_a Q^{2'}_{ai} \nonumber
\\
&   &+ {\displaystyle{\frac{1}{q^3} (1-q^{2})
\sum^{j-1}_{a=i+1}}} \, f^{\dagger}_a Q^{2'}_{ia}, \quad
{\rm for} \quad k = j \nonumber  \\
& = & {\displaystyle{\frac{1}{q^3} f^{\dagger}_i Q^{2'}_{ij} + \frac{1}{q^3}
(1-q^{2}) \sum^{i-1}_{a=1}}} f^{\dagger}_a
Q^{2'}_{aj}, \quad {\rm for} \quad k = i   
\end{array}    \hfill \eqno(34)   $$  

$$\begin{array}[b]{lcl}
Q^{20}_{ij} b^{\dagger}_k &   = & {\displaystyle{\frac{1}{q^2} b^{\dagger}_k Q^{20}_{ij},
\quad {\rm for} \quad k \neq i \quad {\rm or} \quad j}} \nonumber \\
&   = & {\displaystyle{\frac{1}{q^3} b^{\dagger}_j Q^{20}_{ij} + \frac{1}{q^5}
(1-q^{4}) b^{\dagger}_i Q^{21}_{ii} +\frac{1}{q^3}
(1-{q^2} ) \, \sum^{j-1}_{a=i+1}}} \,
b^{\dagger}_a \, Q^{20}_{ia} \nonumber  \\
&     &  +{\displaystyle{\frac{1}{q^4} (1-{q^2} ) \,
\sum^{i-1}_{a=1}}} \, b^{\dagger}_a Q^{20}_{ai}, \quad {\rm
for} \quad  k = j \nonumber  \\
&   = & {\displaystyle{\frac{1}{q^3} b^{\dagger}_i Q^{20}_{ij} + \frac{1}{q^3}
(1-{q^2} ) \, \sum^{i-1}_{a=1}}} \,
b^{\dagger}_a \, Q^{20}_{aj}, \quad {\rm for} \quad  k = i  
\end{array}   \hfill  \eqno(35)   $$ 

$$\begin{array}[b]{lcl}
Q^{21}_{ii} b^{\dagger}_k & = & {\displaystyle{\frac{1}{q^2} b^{\dagger}_k Q^{21}_{ii},
\quad {\rm for} \quad  k \neq i}}  \nonumber \\
& = & {\displaystyle{\frac{1}{q^4} b^{\dagger}_i Q^{21}_{ii} + \frac{1}{q^3} (1-{q^2}) 
 \, \sum^{i-1}_{a=1}}} \, b^{\dagger}_a Q^{20}_{ai},
\quad {\rm for} \quad  k=i   
\end{array}   \hfill    \eqno(36)    $$   

$$  \begin{array}{lcl} 
Q^{22}_{ii} b^{\dagger}_k  & = &  b^{\dagger}_k Q^{22}_{ii}, \quad {\rm
for} \quad k \neq i   \\
& = & b^{\dagger}_i Q^{22}_{ii} + (1 -q^2)
{\displaystyle{\sum^{i-1}_{a=1} b^{\dagger}_a Q^{24}_{ai} +
\frac{1}{q} (q^2 - 1) \sum^{n}_{a=i+1}}} b^{\dagger}_a
Q^{24}_{ia}, \quad {\rm for} \quad k = i       
\end{array}  \hfill \eqno(37)
$$

$$ \begin{array}[b]{lcl}
Q^{23}_{ij} b^{\dagger}_k & = & b^{\dagger}_k Q^{23}_{ij}, \quad {\rm
for} \quad k \neq i \quad {\rm or} \quad j \nonumber \\
& = & {\displaystyle{\frac{1}{q} b^{\dagger}_j Q^{23}_{ij} + 
\frac{1}{q}(1-q^{2}) \sum^{i-1}_{a=1}}} b^{\dagger}_a Q^{24}_{ai} + 
{\displaystyle{\frac{1}{q}(1-q^2) b^\dagger_i Q^{22}_{ii}}} \nonumber \\
&   & {\displaystyle{+ \frac{1}{q}(1-q^2) \, \sum^{k-1}_{a=i+1}}} \,
b^{\dagger}_a Q^{23}_{ia}, \quad {\rm for} \quad k = j \nonumber \\
& = & {\displaystyle{qb^\dagger_i Q^{23}_{ij} - \frac{1}{q^2} (q^2-1)^2
\sum^{i-1}_{a=1} b^\dagger_a Q^{24}_{aj} + \frac{1}{q^2} (q^2 -1) 
b^\dagger_j Q^{22}_{ii}}} \\
& + & {\displaystyle{\frac{1}{q}(q^{2}-1) \sum^{j-1}_{a=i+1} 
b^\dagger_a Q^{23}_{ja} + \frac{1}{q}(q^{2}-1) \sum^n_{a=j+1} 
b^\dagger_a Q^{24}_{ja}}} \ \,, \ {\rm for} \quad k = i   
\end{array}  \hfill \eqno(38)
$$

$$ \begin{array}[b]{lcl}
Q^{24}_{ij} b^{\dagger}_k & = & b^{\dagger}_k Q^{24}_{ij},\quad {\rm
for} \quad k \neq i \quad {\rm or}\quad j \nonumber  \\
& = & q b^{\dagger}_j Q^{24}_{ij} + {\displaystyle{\frac{1}{q}(q^{2}-1) \,
\sum^{n}_{a=j+1} b^{\dagger}_a Q^{24}_{ia}, \quad {\rm
for} \quad k = j}} \nonumber  \\
& = & {\displaystyle{\frac{1}{q} b^{\dagger}_i Q^{24}_{ij} +
\frac{1}{q}(1-q^{2}) \, \sum^{i-1}_{a=1}}} b^{\dagger}_a Q^{24}_{aj}, 
\quad {\rm for} \quad k = i  \end{array}   \hfill   \eqno(39)
$$

We see that all the above equations are of the form (30).  This is not
the complete set of consistency conditions; to get these we will need
the commutation properties between $f$ and $f^{\dagger}$ which are yet
to be obtained.  However, the validity of the consistency conditions
(33-39) already points to the existence of an underlying Fock space and
encourages us to find it.

\vskip .5cm

\noindent{\bf 6. Completion of the Fock space realization}

We find the required Fock space by showing that the ${b,f}$ system is in
fact related to the canonical bose-fermi system ${\tilde{b}, \tilde{f}}$
defined by the usual algebra (for all $i$ and $k$):

$$
[{\tilde{b}}_i, {\tilde{b}}^{\dagger}_k] =
\delta _ {i k }\quad ; \quad [{\tilde{b}_i}, {\tilde{b}_k}] = 0
\eqno(40)
$$
$$
\left\{\tilde{f}_i, \tilde{f}^{\dagger}_k\right\} =
\delta _{i k} \quad ; \quad \left\{\tilde{f}_i, \tilde{f}_k\right\} = 0
\eqno(41)
$$
$$
[{\tilde{b}}_i, {\tilde{f}}^{\dagger}_k] =
0 \quad ; \quad [{\tilde{b}}_i {\tilde{f}}_k] = 0
\eqno(42)
$$
\noindent and their hermitian conjugates where $[x,y]$ and $\{x,y\}$
denote the usual commutator and anticommutator respectively.  The relationship is given by
the transformation equations :
$$
b^{\dagger}_i = q^{\sum_{k > i} \tilde{N}_k} \left(
\frac{[\tilde{N}_i]}{\tilde
{N}_i} \right)^{1/2} \, \tilde{b}^{\dagger}_i
\eqno(43)
$$
$$
f^{\dagger}_i  =  q^{\sum_{p < i} \tilde{M}_p - \Sigma_p \tilde{N}_p -
\tilde{N}_i} \, \tilde{f}^{\dagger}_i + (1-q^2) {\displaystyle{\sum_{k <
i}}} 
q^{\sum_{p<k} \tilde{M}_p - \Sigma_p \tilde{N}_p - \sum_{k \leq p
\leq i} \tilde{N}_p} \, \left( \frac{[\tilde{N}_k + 1][\tilde{N}_i]}{(\tilde{N}_k + 1)(\tilde{N}_i)} \right)^{1/2} \, \tilde{b}^{\dagger}_i \tilde{b}_k
\tilde{f}^{\dagger}_k  \eqno(44) 
$$
\noindent and their hermitian conjugates.  The operators $\tilde{N}_i$
and $\tilde{M}_i$ are the number operators of the ${\tilde{b},
\tilde{f}}$ system :
$$
\tilde{N}_i = \tilde{b}^{\dagger}_i \tilde{b}_i \quad ; \quad \tilde{M}_i =
\tilde{f}^{\dagger}_i \tilde{f}_i    \eqno(45)
$$

\noindent and they satisfy the usual commutation relations: 
$$
[{\tilde{N}}_i, {\tilde{N}}_k] = [{\tilde{M}}_i, {\tilde{M}}_k] =
[{\tilde{N}}_i, {\tilde{M}}_k ] = 0.  \eqno(46) 
$$
$$
[ \tilde{N}_i, \tilde{b}^{\dagger}_k] = \delta_{ik}
\tilde{b}^{\dagger}_k \quad ; \quad [\tilde{M}_i, \tilde{b}^{\dagger}_k] = 0
\eqno(47)
$$
$$
[ \tilde{M}_i, \tilde{f}^{\dagger}_k] = \delta_{ik}
\tilde{f}^{\dagger}_k \quad ; \quad [\tilde{N}_i, \tilde{f}^{\dagger}_k] = 0
\eqno(48)
$$
\noindent In (43) and (44) the square bracket enclosing a single object
[L] is defined by
$$
[L] = \frac{q^{2L} - 1}{q^2-1} = 1+ q^2 + q^4 + \ldots q^{2(L-1)}
\eqno(49)
$$
\noindent Using (43) and (44), the algebra given by 
(1$^{\prime}$-13$^{\prime}$, 20-27)
can be verified by straightforward but long computations.

The transformations (43) and (44) can be inverted (for $q \neq 0$) to
give 
$$
\tilde{b}^{\dagger}_i = q^{- \sum_{k > i} \tilde{N}_k} \, \left(
\frac{\tilde{N}_i}{[\tilde{N}_i]} \right)^{1/2} b^{\dagger}_i
\eqno(50)
$$

$$
\tilde{f}^{\dagger}_i  =  q^{- \sum_{p < i} \tilde{M}_p + \sum_p
{\tilde{N}}_p + {\tilde{N}}_i} \, f^{\dagger}_i + q^{2i} (q^2 - 1) 
q^{-
\sum_{p < i} \tilde{M}_p + \sum_{p < i} \tilde{N}_p -\sum_{p > i}
\tilde{N}_p}\quad {\displaystyle{\sum_{k < i}}} q^{-2k} b^{\dagger}
_i b_k
f^{\dagger}_k    \eqno(51)  
$$

The transformation given by (43) and (44) is our central result.  This
establishes the complete Fock space of the ${b,f}$ system, since the
latter has been expressed in terms of the ${\tilde{b}, \tilde{f}}$
system which operates on the canonical Fock space of bosons and
fermions.

Nevertheless, one might still want to know the q-commutation relations
between $f$ and $f^{\dagger}$.  These are now derivable from (43) and
(44).  After a long computation we get

$$
f_i f^{\dagger}_j + q f^{\dagger}_j f_i = q (1-q^2) b^{\dagger}_j b_i
A_{ij} \quad {\rm for} \quad i < j     \eqno(52)
$$

$$
f_i f^{\dagger}_i + f^{\dagger}_i f_i  =  B_i - (1-q^2)^2
{\displaystyle{\sum_{\stackrel{k < i,}{k \,\neq}}}}
{\displaystyle{\sum_{\stackrel{k' < i}{k'}}}} \, \tilde{f}^{\dagger}_k
\tilde{f}_{k'} \tilde{b}^{\dagger}_{k'} \, \tilde{b}_{k} C_{i k k'}
\eqno(53)
$$
\noindent where A,B and C are functions of the number operators
${\tilde{N}}_i$ and ${\tilde{M}}_i$:


$$
A_{ij}   =  \left( q^{-2 \sum_{k \ge i} \tilde{N}_k} + 
(1-q^2) {\displaystyle{\sum_{k < i}}} [{\tilde{N}}_k + {\tilde{M}}_k] \
q^{-2 \sum_{p \geq k} \tilde{N}_p} \right) q^{2 {\sum_{p < i}}
\tilde{M}_p - 2 \sum_p \tilde{N}_p}  
\eqno(54)
$$

$$ \begin{array}{lcl}
B_i &  =&  q^{2 \sum_{p < i} \tilde{M}_p - 2 \sum_p
\tilde{N}_p - 2 \tilde{N}_i} \\
\\
& +& {\displaystyle{\sum_{k < i} (1-q^2)^2 \left\{ [\tilde{N}_k]
[\tilde{N}_i + 1] - (1-q^2)^{-1} \tilde{M}_k \left( q^{2 \tilde{N}_i} - q^{2
\tilde{N}_k} \right) \right\} q^{2 \sum_{p < k} \tilde{M}_p - 2
\sum_{k \leq p \leq i} \tilde{N}_p -2 \sum_{p} \tilde{N}_p}}} 
\end{array} \eqno(55) $$ 

$$
C_{i k k'}  =  \left( \frac{[\tilde{N}_k][\tilde{N}_{k'} +1]}{\tilde{N}_k (\tilde{N}_{k'} +1)}
\right)^{1/2} q^{2 \tilde{N}_i + (\sum_{p < k'} + \sum_{p < k}) \tilde{M}_p - (\sum_p +
\sum_{k \leq p \leq i} + 
\sum_{k'\leq p \leq i})\tilde{N}_p} 
\eqno(56)
$$

\noindent The right hand side of (53) contains the canonical operators
$\tilde{b}, \tilde{f}$ and these can be reexpressed in terms of $b,f$
using the inverse relations (50)  and (51); we have not written them
in that form since the expressions would be longer.

Further, one may like to specify the commutation relations of
$\tilde{N}_i$ and $\tilde{M}_i$ with respect to $b, f, b^{\dagger}$ and
$f^{\dagger}$ since these will be required for the closure of the
algebra.  These are (for all $i$ and $k$)
$$
[\tilde{N}_i, b^{\dagger}_k] = \delta_{ik} b^{\dagger}_k     \eqno(57)
$$
$$
[\tilde{M}_i, b^{\dagger}_k] = 0      \eqno(58)
$$
$$ 
[\tilde{M}_i, f^{\dagger}_k ]  =  \delta_{ik} f^{\dagger}_k -
(1-q^2) {\displaystyle{\sum_m}} D_{imk} b^{\dagger}_k b_m f^{\dagger}_m
\eqno(59)
$$
$$
[\tilde{N}_i, f^{\dagger}_k] = (1-q^2) 
{\displaystyle{\sum_m}} D_{imk} b^{\dagger}_k b_m f^{\dagger}_m 
\eqno(60)
$$
\noindent together with the hermitian conjugates of these relations,
where

$$ \begin{array}{lcl }
D_{ikm} &  =  & \delta_{ik} \theta_{mk} q^{2(k-m)-2 \sum_{p \ge k}
{\tilde{N}}_p} \\
\\
& & \quad - \theta_{ik} \left\{ \delta_{mi} \, q^{2 - 2 \sum_{p > i}
\tilde{N}_p} - \theta_{mi} (1-q^{2(\tilde{N_i} + 1)}) q^{2(i-m) -2
\sum_{p \geq i} \tilde{N}_p} \right\}  \end{array}  \eqno(61)
$$
In eq.(61), $\theta_{ik}$ is defined to be 1 for $i < k$ and zero
otherwise.

Although we have given the $f f^{\dagger}$ relations (52,53)
supplemented by (57-61) for the sake of exhibiting the complete
algebra of ${b,f}$ system and thus completing the Fock space
representation of the differential calculus on the noncommuting quantum
space, the alternative way of expressing our result in terms of the
transformation equations (43) and (44) is simpler and more
transparent.

In particular, the  origin of the transmutations discussed in Sec.4 becomes clear now.
Eq.(44) shows that the fermionic operator $f^{\dagger}_i$ 
creates not only fermion $\tilde{f}_i$ but also the boson
$\tilde{b}_i$, at the same time converting the boson $\tilde{b}_k$ into
fermion $\tilde{f}_k$ for all $k < i$.

Relations of the type (43) which can be called generalized
Klein-Jordan-Wigner relations are known from the earlier literature$^{11,12}$
but the relation (44) which leads to the idea of transmutation
is new.

\vskip .5cm

\noindent{\bf 7. Discussion}

We have constructed here the complete Fock space associated with the
differential calculus on the quantum space.  The algebraic relations
between the creation and annihilation operators spanning the Fock space
have been derived and their internal consistency established.  

The present formalism leads to the notion of statistical transmutation
between different kinds of quanta residing in the generalized Fock
space.  Consequently, the number operators for individual quanta are not
conserved, only certain partial sums of number operators are conserved.

We have been able to map the entire set of new creation and annihilation
operators to the creation and annihilation operators for canonical
fermions and bosons.  Because of the existence of such transformations,
we can say that as far as the underlying Fock space is concerned, the
deformations leading to covariant differential calculus do not lead to
anything fundamentally new.  What one gets is only a different avatar of
the canonical algebra of fermions and bosons$^{10}$, however with
statistical transmutation.  Thus the formalism presented here demystifies
the non-commutative differential calculus on which so much recent work
has been done, by providing its representation in a linear vector space
which is a composite of canonical fermionic and bosonic spaces.

The insight gained through this understanding of the noncommutative
differential calculus in terms of Fock space may prove useful for
further lines of investigation.  Some of these are : \\

\noindent (a) Is it possible to have a nontrivial deformation of the
differential calculus that does not require the statistical
transmutation ?  This may require the dropping of the Leibnitz rule
(15).
\vskip .5cm
\noindent (b) More general transmutations can be introduced if we
replace (43) and (44) by
$$
b^{\dagger}_i = F_i \tilde{b}^{\dagger}_i + {\displaystyle{\sum_j}}
G_{ij} \tilde{f}^{\dagger}_i \tilde{f}_j \, \tilde{b}^{\dagger}_j
\eqno(62)
$$
$$
f^{\dagger}_i = H_i \tilde{f}^{\dagger}_i + {\displaystyle{\sum_j}}
K_{ij} \tilde{b}^{\dagger}_i \tilde{b}_j \, \tilde{f}^{\dagger}_j
\eqno(63)
$$

\noindent where $\tilde{b}_i$ and $\tilde{f}_i$ are canonical boson and
fermion annihilation operators and $F_i, G_{ij}, H_i$ and $K_{ij}$ are
functions of the canonical number operators as well as one or more
deformation parameters.  Hence, the path is open, to construct a 
variety of new deformed differential calculi.  One may also remark 
that transformations of the
type (62) and (63) can be used to describe the transmutation of any two
species of canonical quanta ; both may be bosons or fermions.

Differential calculus is generated by three operators $(x,
\frac{\partial}{\partial x}, dx)$, whereas Fock space is spanned by
$(b^{\dagger}, b, f^{\dagger},f)$.  What is the significance of the
additional operator $f,$ with regard to differential calculus?  A
possible answer may be provided in the context of the Lagrangian and
Hamiltonian dynamics in quantum space$^{14}$ in which the
$q$-deformed differential calculus plays an essential role.  One may
introduce the velocity $\dot{x}$ as the differential $dx$ divided by
$dt$ where $t$, the time, is taken to be a commuting number.  The
dynamical formalism will require the calculus to be extended to the
derivative $\frac{\partial}{\partial {\dot{x}}}$.  For this extended
calculus, one may have the mapping :
$$
x \rightarrow b^{\dagger} \quad ; \quad \frac{\partial}{\partial x}
\rightarrow b \quad ; \quad \dot{x} \rightarrow f^{\dagger} \quad ;
\quad \frac{\partial}{\partial \dot{x}} \rightarrow f.   \eqno(64)
$$

Finally we note that the transformations linking the new operators to
canonical operators become ill-defined when $q \rightarrow 0$ \, or \,
$\pm \infty$.  For such singular values of $q$, new kinds of statistics
(`null statistics') living in new Fock spaces (`Fock spaces of frozen
order') emerge$^{13}$.  Statistical transmutation can be incorporated
into the null statistics too, leading to newer structures.

\end{document}